\documentclass[showpacs,superscriptaddress,amsmath,amssymb,twocolumn]{revtex4}
\usepackage{graphicx}
\usepackage{color}
\usepackage{bm}

\begin{document}
\title[DI of BEC with HOI in OL]{Dynamical instability of a Bose-Einstein condensate with higher-order interactions in an optical potential through a variational approach}

\author{E. Wamba}
\affiliation{Department of Physics, Faculty of Science, University
of Yaound\'{e} I, P.O. Box 812, Yaound\'{e}, Cameroon}
\affiliation{African Institute for Mathematical Sciences, 6 Melrose
Road, Muizenberg, 7945, South Africa}
\author{S. Sabari}
\affiliation{Department of Physics, Pondicherry University,
Puducherry 605014, India}
\author{K. Porsezian}
\affiliation{Department of Physics, Pondicherry University,
Puducherry 605014, India}
\author{A. Mohamadou}
\affiliation{Department of Physics, Faculty of Science, University
of Douala, P.O. Box 24157, Douala, Cameroon}
\affiliation{The Abdus Salam International Centre for Theoretical
Physics, P.O. Box 586, Strada Costiera 11, I-34014, Trieste, Italy}
\author{T.C. Kofan\'{e}}
\affiliation{Department of Physics, Faculty of Science, University
of Yaound\'{e} I, P.O. Box 812, Yaound\'{e}, Cameroon}

\begin{abstract}
We investigate the dynamical instability of Bose-Einstein
condensates (BECs) with higher-order interactions immersed in an
optical lattice with weak driving harmonic potential. For this, we
compute both analytically and numerically a modified
Gross-Pitaevskii equation with higher-order nonlinearity and
external potentials generated by magnetic and optical fields.
Using the time-dependent variational approach, we derive the
ordinary differential equations for the time evolution of the
amplitude and phase of modulational perturbation. Through an
effective potential, we obtain the modulational instability
condition of BECs and discuss the effect of the higher-order
interaction in the dynamics of the condensates in presence of
optical potential. We perform direct numerical simulations to
support our analytical results, and good agreement is found.

\end{abstract}

\maketitle

\section{Introduction}

Over the past two decades, the dynamics and properties of
Bose-Einstein condensates (BECs) in optical lattices (OLs) have
been intensively investigated both experimentally \cite
{Anderson1998,Hagley1999,Eiermann2004} and theoretically
\cite{Berg-Sorensen1998,Choi1999,Malomed1999,Porter2004}. In these
works, the OL created by the interference of laser beams induces a
periodic potential that traps the bosonic atoms. The properties of
the atoms are characterized by the depth and period of the
optically induced potential. An interesting feature of this
lattice potential is that its intensity can be modulated from very
weak to very strong \cite{Anderson1998}. BECs in periodic
potentials have been used to investigate many physical phenomena,
such as Josephson effect \cite{Cataliotti2001}, Bloch oscillations
\cite{Anderson1998,Morsch2001,Cristiani2002}, Landau-Zener
tunneling \cite{Wu2003}, solitons \cite{Eiermann2003}, quantum
phase transitions of the Mott insulator type \cite{Greiner2002},
superfluid and dissipative dynamics \cite{Wu200203}, phase diagram
\cite{Zheng2005}, and nonlinear dynamics of a dipolar
\cite{Xie2005} or spinor BEC \cite{Li2005}.

Most findings of experiments in BECs are reproduced and described by
the theoretical model based on the nonlinear mean-field
Gross-Pitaevskii (GP) equation \cite{Dalfovo1999}. The nonlinear
term arises in the GP equation due to the effect of the inter-atomic
interaction in the condensate which is described by the
\emph{s}-wave scattering length, $a_s$.
The interaction strength can be controlled by using different
experimental techniques.
Notably, the strength and sign of the atomic scattering length can
be varied by tuning the external magnetic field near Feshbach
resonances \cite{Feshbach}.
When the sign is positive, the interaction in the BEC is
repulsive. In the presence of OL, such interaction can give rise
to stable localized matter-wave states in the form of gap
solitons.
The BEC gap soliton was predicted theoretically
\cite{Abdullaev2001,Carusotto2002,Louis2003} and demonstrated
experimentally \cite{Eiermann2004}. Gap solitons are represented
by stationary solutions to the respective GP equation, with the
eigenvalue (chemical potential) located in a finite bandgap of the
OL-induced spectrum \cite{Morsch2006}. In BECs with attractive
interactions $(a_{s} < 0)$, solitons realize the ground state of
the condensate. Such solitons were created in condensates of
$^7$Li \cite{Strecker2002} and $^{85}$Rb \cite{Cornish2006} atoms,
with the sign of the atomic interactions switched to attraction by
means of the Feshbach-resonance technique. In the presence of a
periodic potential, such solitons should exist too as first shown
in the context of optical setting \cite{Malomed1999}, and later
demonstrated in detail in the framework of BECs through a GP
equation \cite{Alfimov2002a,Salasnich2007}.
However, the simple mean-field GP equation is less convenient in
some contexts. Higher-order terms in the expansion of the phase
shifts at low momenta, determined by the effective range, the
shape parameter etc., give corrections to the simple GP equation.
It has been shown that the critical number of condensed atoms
needed for stability, the chemical potential, the condensate
profiles and the energy levels of a harmonically trapped BEC
strongly depend on the higher-order scattering term when the
scattering length approaches zero  \cite{Zinner2009}.
Furthermore, because the higher-order interactions is determined
by the shape parameter we can expect that the higher-order
interactions would strongly be affected by the trap potential.
Then studying the dynamics of BECs considering the presence of
higher-order interactions becomes relevant especially in the case
of narrow resonances.

In many dispersive systems described by nonlinear wave equations,
it appears the general phenomenon of modulational instability
(MI). For nonlinear systems in periodic potentials, the MI is
usually referred to as the \emph{dynamical instability} since it
is directly connected to the dynamic equation which describes the
system \cite{Desarlo}. It occurs when the eigenspectrum of the
excitations of the system exhibits complex frequencies. In this
case, due to the interplay between nonlinearity, dispersion and
periodicity, arbitrary small perturbations of the wave function
may grow exponentially, eventually leading to the destruction of
the initial state
\cite{Desarlo,Agrawal2001,Konotop2002,Fallani2004,Barontini}. MI
has been demonstrated both experimentally
\cite{Fallani2004,miexp1,miexp2,miexp3} and theoretically
\cite{mitheo1,mitheo2,mitheo6}. Recently, many investigations have
been devoted to the MI of both single-component BECs and
double-species BECs in optical lattices
\cite{Desarlo,Huang2010,Rapti2004,Fallani2004,Barontini,Ruostekoski2007}.
Moreover, the studies with relation to the MI have also attracted
much interest, as the MI is an indispensable mechanism for
understanding the relevant dynamic processes in the BEC systems,
which include domain formation \cite{Kasamatsu2004,Zhang2005},
generation and propagation of solitonic waves
\cite{Kevrekidis2003,Kasamatsu2006} and quantum phase transition
\cite{Smerzi2002}, etc.
The study of MI is usually conducted by means of the standard
linear stability analysis, which unfortunately cannot help
efficiently in the context of OL.
In this paper we reexamine the MI in the nonlinear Schr\"{o}dinger
(NLS) equation with the effect of periodic potential and
higher-order nonlinearity with the help of time-dependent
variational approach and numerical calculations.

The paper is organized as follows: in Sec.~\ref{sec2}, we present
the theoretical model that describes the condensates with
higher-order nonlinearity in OL. Sec.~\ref{sec3} is devoted to the
analytical framework in which we derive the MI conditions of the
system through the time-dependent variational approach
and predict the dynamics of the system in both cases of attractive
and repulsive interactions.
Then, in Sec.~\ref{sec4}, we perform direct numerical integrations
to check the validity of the MI conditions found by analytical
methods and study the interplay between higher-order nonlinearity
and optical potential. Sec.~\ref{sec5} summarizes our results and
conclude the work.

\section{Theoretical model}\label{sec2}

\textbf{In the ultracold regime where the temperature is much
smaller than the critical temperature for condensation, a Bose gas
may obey the $T = 0$ formalism. The higher-order effects in the
two-body scattering dynamics can be captured by an energy
functional \cite{fabrocini,fu} which allows the derivation of the
following generalized (or modified)} GP equation
\cite{collin,ripoll}
\begin{align}
\begin{split}
\mathrm{i}\hbar\frac{\partial \psi(\mathbf{r},t)}{\partial t} = &\,-\frac{\hbar^2}{2m}\nabla^2\psi(\mathbf{r},t) + V_{\mathrm{ext}}(\mathbf{r})\psi(\mathbf{r},t)\\
+ &\, g_0 |\psi(\mathbf{r},t)|^2 \psi(\mathbf{r},t) + \eta
\nabla^2 (|\psi(\mathbf{r},t)|^2) \psi(\mathbf{r},t).
\label{Jaco1}
\end{split}
\end{align}
In Eq. (\ref{Jaco1}) $\hbar$ is the reduced Planck's constant, $m$
is the mass of the boson, and $g_0$ is the strength of the
two-body inter-atomic interactions defined by $g_0=4\pi\hbar^2
a_s/m$, with $a_s$ being the \emph{s}-wave scattering length.
\textbf{The strange nonlinear term which depends on the Laplacian
of the number density is the so-called \emph{residual
nonlinearity} \cite{ripoll}. It describes the shape-dependent
confinement correction of the two-body collision potential. The
parameter $\eta$ is the higher-order scattering coefficient which
depends on both the \emph{s}-wave scattering length and the
effective range for collisions \cite{fabrocini,fu,collin}. This
parameter reads $\eta=g_0 g_2$, where $g_2$ is defined by
$g_2=a_s^2 /3-a_s r_e /2$, with $r_e$ being effective range.
The model in Eq. (\ref{Jaco1}) was derived and explained in
\cite{collin}(and Refs. therein) for any general external
potential $V_{\mathrm{ext}}(\mathbf{r})$, and the physical meaning
of the residual nonlinearity was discussed in full detail. When
the Bose gas with higher-order interaction (HOI) is immersed in a
trap consisting of an OL driven by a highly elongated harmonic
trap \cite{Konotop2002,Fallani2004}, the dynamics of the
condensate is governed by the following modified GP equation}
\begin{align}
\begin{split}
\mathrm{i}\hbar\frac{\partial \psi(\mathbf{r},t)}{\partial t} = &\,-\frac{\hbar^2}{2m}\nabla^2\psi(\mathbf{r},t)+ \frac{m}{2}(\omega^2_{\perp} \rho^2+\omega^2_x x^2)\psi(\mathbf{r},t)\\
&\, + V_{\mathrm{OL}}(x)\psi(\mathbf{r},t) + g_0 |\psi(\mathbf{r},t)|^2 \psi(\mathbf{r},t)\\
&\, + \eta \nabla^2 |\psi(\mathbf{r},t)|^2 \psi(\mathbf{r},t),
\label{Jac1}
\end{split}
\end{align}
where $\omega_{\perp}$ and $\omega_x$, respectively, are the
radial and longitudinal frequencies of the anisotropic trap
($\omega_{\perp} \neq \omega_x$), and $\rho=\sqrt{y^{2}+z^{2}}$
denotes the radial distance. The OL potential is applied only in
the axial direction, such as to have $V_{\mathrm{OL}}(x)= V_{max}
\cos^2(\kappa x)$, with $V_{max}$ and $\kappa$ being the effective
depth and the wave number of the optical potential, respectively.

In the case of elongated or cigar-shaped condensate
($\omega_{\perp} \gg \omega_x$), we can
make the change $\psi(\mathbf{r},t) = \phi_0(\rho)\phi(x,t)$, where
$\phi_0=\sqrt{\frac{1}{\pi
a^2_{\perp}}}\exp(-\frac{\rho^2}{2a^2_{\perp}})$ with
$a_{\perp}=\sqrt{\hbar/m\omega_{\perp}}$, is the ground state of the
radial equation
\begin{align}
\begin{split}
-\frac{\hbar^2}{2m}\nabla^2_\rho \phi_0+ \frac{m}{2}\omega^2_{\perp}
\rho^2 \phi_0 = \hbar \omega_{\perp}\phi_0. \label{Jac2}
\end{split}
\end{align}
Then, multiplying both sides of the GP Eq.~(\ref{Jac1}) by
$\phi^*_0$ and integrating over the radial variable $\rho$, we
obtain a quasi-1D GP equation that reads:
\begin{align}
\begin{split}
\mathrm{i}\hbar\frac{\partial \phi(x,t)}{\partial t} = &\,\left[-\frac{\hbar^2}{2m} \nabla_x^2 + \left(\frac{m}{2}\omega^2_x x^2+ V_{max} \cos^2(\kappa x)\right)\right]\phi \\
&\, +\left[\frac{g_0}{2\pi a^2_{\perp}} |\phi|^2 +\frac{\eta}{2\pi
a^2_{\perp}} \nabla_x^2 |\phi|^2 \right]\phi. \label{Jac3}
\end{split}
\end{align}
In most cases, the above Eq.~(\ref{Jac3}) is used in a
dimensionless form \cite{FKhAbdul2005,Zhang2007,Zhang2003}. For
this purpose we need to introduce the following change of
variables: $t \sim t \nu$, $x \sim x \kappa$, $\phi \sim \phi
\sqrt{2a_{s0}\omega_{\perp}/\nu}$, where $\nu=E_R/\hbar$,
$\alpha=\omega^2_x/4\nu^2$ and $E_R=\hbar^2\kappa^2/2m$. Then, we
come to the following normalized 1D GP equation with harmonic and
optical potentials:
\begin{align}
\begin{split}
\mathrm{i}\frac{\partial \phi(x,t)}{\partial t} = &\,-\nabla_x^2
\phi+ (\alpha x^2+V_s
\cos^2(x))\phi + \\
&\, g|\phi|^2 \phi+g \sigma \nabla_x^2 |\phi|^2 \phi, \label{Jac4}
\end{split}
\end{align}
where $ \nabla_x^2=\frac{\partial^2}{\partial x^2}$,
$V_s=V_{max}/E_R$, $\sigma=g_2\kappa^2$, $a_s=g\, a_{s0}$, with
$a_{s0}$ being the constant scattering length, $g=\pm{1}$ is the
sign of the scattering length. Some special cases of that equation
have been considered in previous works \cite{PLA2013,PLA2013b}. In
order to get the analytical condition that may allow to examine
the MI of BECs in the system, we first need to find for Eq.
(\ref{Jac4}) an expression with less space-dependent coefficients.
We begin with a modified lens-type transformation which consists
in setting \cite{Wamba2008a}
\begin{align}
\begin{split}
\phi(x,t) =
&\,\frac{1}{\sqrt{l(t)}}\widetilde{\psi}(\tilde{x},\tilde{t})\exp(\mathrm{i}f(t)x^2).
\label{Jac5}
\end{split}
\end{align}
In this expression, we choose $l(t)=|\cos(2\sqrt{\alpha}t)|$,
$\tilde{x}=\frac{x}{l(t)}$,
$\tilde{t}(t)=\frac{1}{2\sqrt{\alpha}}\tan(2\sqrt{\alpha}t)$, and
$f(t)=-\frac{\sqrt{\alpha}}{2}\tan(2\sqrt{\alpha}t)$. The
rescaling signals the existence of negative $\tilde{t}$ and is
valid for any $t\neq\frac{(2n+1)\pi}{4\sqrt{\alpha}}$ (where $n$
is a positive integer) in the $t$-$\tilde{t}$ plane. We consider
the case where $t$ goes from zero to $\pi/(4\sqrt{\alpha})$ to
ensure a continuous variation of $\tilde{t}$ from zero to
infinity. Let us note in passing that the length scale $l(t)$ can
be written in terms of the new time as
$l(\tilde{t})=1/\sqrt{1+4\alpha \tilde{t}^2}$. For simplicity, we
will drop the tildes in what follows. Then Eq. (\ref{Jac4}), in
terms of the new variables $x$ and $t$, is reduced to

\begin{align}
\begin{split}
\mathrm{i}\frac{\partial \psi(x,t)}{\partial t} = -\nabla_x^2 \psi(x,t)+ V_0(t) \cos^2(x)\psi(x,t) + \\
s(t) |\psi(x,t)|^2 \psi(x,t)+\eta(t)\nabla_x^2 |\psi(x,t)|^2\psi(x,t), \label{Jac6}
\end{split}
\end{align}
where $V_0(t)=V_s \, (1+4\alpha t^2)^{-1}$, $s(t)=g\, (1+4\alpha
t^2)^{-1/2}$ and $\eta(t)=\sigma\, g\, (1+4\alpha t^2)^{1/2}$
are the new rescaled strengths of OL, two-body and HOIs. The 1D GP
Eq. (\ref{Jac6}) can be more easily handled through variational
methods.

\section{Analytical results}\label{sec3}
\subsection{Variational approximation and instability criteria}\label{sec31}

For many physical systems, the variational approximation method
has been found to be an indispensable tool in the investigation of
dynamical properties. Here, we use a time-dependent variational
approach to examine the MI of BECs with HOI trapped in optical
potential. As applied in many works \cite{Wamba2008,sabari2010},
the first step of the process consists in finding the Lagrangian
density that may generate the governing Eq. (\ref{Jac6}). Such
density reads
\begin{align}
\begin{split}
\mathcal{L} =
&\,\frac{\mathrm{i}}{2}\left(\frac{\partial\psi}{\partial
t}\psi^*-\frac{\partial\psi^*}{\partial
t}\psi\right)-|\nabla_x\psi|^2- V_0(t)
\cos^2( x)|\psi|^2 \\
&\,-\frac{s(t)}{2}|\psi|^4
-\frac{\eta(t)}{2}|\psi|^2\nabla_x^2(|\psi|^2). \label{Jac7}
\end{split}
\end{align}
It is easy to show that an exact solution of the NLS Eq.
(\ref{Jac6}) in the absence of optical potential is a wave of the
form
\begin{align}
\begin{split}
\psi(x,t)=A_0 \exp\Big[\mathrm{i}\Big(k x-k^2 t - A^2_0 \int_0^{t}
s(t')\mathrm{d}t'\Big)\Big]. \label{Jac8}
\end{split}
\end{align}

Then we may use, as variational ansatz for the wavefunction of the
condensate, a MI-motivated trial wave function given by the
following ansatz
\begin{align}
\begin{split}
\psi(x,t)=\Big\{A_0+a_1(t)\exp[\mathrm{i}(q x+b_1(t))]+\\
 a_2(t)\exp[\mathrm{i}(-q x+b_2(t))]\Big\}\\ \times \exp\Big[\mathrm{i}\Big(k x-k^2 t - A^2_0
\int_0^{t} s(t')\mathrm{d}t' \Big)\Big].\label{Jac10}
\end{split}
\end{align}

This trial wave function may be substituted into the Lagrangian
density which may be integrated over the entire space to obtain
the effective Lagrangian. A tricky way of computing the effective
Lagrangian for this type of problem is to consider a circular (1D)
geometry, which imposes periodic boundary conditions on the
wavefunction $\psi(x, t)$ and integration limits $0 \leqslant x <
2\pi$. This causes the quantization of the wave numbers, i.e. $k,
q = 0,\pm1,\pm2,\pm3, . . . $. In this new geometry, calculating
the effective Lagrangian yields

\begin{align}
\begin{split}
L_{\mathrm{eff}} = -\pi  \Big\{A_0^2\left[V_0-A_0^2\,s(t)\right]\\
+a_1^2\left[2q(q+2k-A_0^2\,q\,\eta(t))+ s(t)(2A_0^2+a_1^2)+2\dot{b}_1+V_0\right]\\
+a_2^2\left[2q(q-2k-A_0^2\,q\,\eta(t))+s(t)(2A_0^2+a_2^2) +2\dot{b}_2+V_0\right]\\
+4\,a_1\,a_2\,s(t)\left[a_1a_2+A_0^2\cos(b_1+b_2)\right]\\
-4\,a_1\,a_2\,q^2\,\eta(t)\left[2a_1a_2+A_0^2\cos(b_1+b_2)\right]
\Big\}.\label{Jac12}
\end{split}
\end{align}
The expression of this effective Lagrangian is such that the pair
$\{b_1(t), b_2(t)\}$ may be interpreted as the set of generalized
coordinates of the system, while the pair $\{A_1(t), A_2(t)\}$,
with $A_1(t)=2a_1^2(t)$ and $A_2(t)=2a_2^2(t)$, gives the
corresponding momenta. The Hamiltonian of the system is expressed
as

\begin{align}
\begin{split}
 H = &\, -L+\int_{-\infty}^{\infty}\frac{\mathrm{i}}{2}\left(\frac{\partial\psi}{\partial t}\psi^*-\frac{\partial\psi^*}{\partial t}\psi\right)\mathrm{d}x.\label{Jac13}
\end{split}
\end{align}
Considering the integration limits imposed by the new geometry, we
have
\begin{align}
\begin{split}
H = \pi  \Big\{A_0^2\left[2k^2+V_0+A_0^2 s(t)\right]+\frac{s(t)}{4}\left(A^2_1+A^2_2\right) \\
+A_1\left[(k+q)^2+\frac{V_0}{2}-A_0^2\,q^2\,\eta(t)+2A_0^2s(t)\right]\\
+A_2\left[(k - q)^2+\frac{V_0}{2}-A_0^2\,q^2\,\eta(t)+2A_0^2s(t)\right]\\
+\sqrt{A_1}\,\sqrt{A_2}\,s(t)\left[ 2\,A_0^2\,\cos(b_1+b_2)+\sqrt{A_1}\,\sqrt{A_2}\right]\\
-\sqrt{A_1}\,\sqrt{A_2}\,q^2\,\eta(t)\left[
2\,A_0^2\,\cos(b_1+b_2)+\sqrt{A_1}\,\sqrt{A_2}\right]\Big\},\label{Jac14}
\end{split}
\end{align}

In order to derive the evolution equations for the time-dependent
parameters introduced in Eq.~(\ref{Jac10}), we use the
corresponding Euler-Lagrange equations based on the variational
effective Lagrangian $L_{\mathrm{eff}}$. In the generalized form,
these equations read
\begin{align}
\begin{split}
 \frac{\mathrm{d}}{\mathrm{d}t} \left( \frac{\partial  L_{\mathrm{eff}}}{\partial \dot \xi_i}\right)-\frac{\partial  L_{\mathrm{eff}}}{\partial \xi_i}  = 0 \label{Jac15},
\end{split}
\end{align}
where $\xi_i$ and $\dot{\xi_i}$ are, respectively, the generalized
coordinate and corresponding generalized momentum. Hence, the
evolution equation corresponding to the variational parameter $a_1$
is
\begin{align}
\begin{split}
\frac{\partial a_1}{\partial t} =
&\,\left(q^2\,\eta(t)-s(t)\right)A_0^2\,a_2\,\sin(b_1+b_2).\label{Jac16}
\end{split}
\end{align}
For the parameter $b_1$ the evolution equation reads
\begin{align}
\begin{split}
\frac{\partial b_1}{\partial t} =
-2kq-q^2-\frac{V_0}{2}-s(t)\left(A_0^2+a_1^2+2a_2^2\right)+
q^2\,\eta(t)\\
\times\left(A_0^2+4a_2^2\right)+\left(q^2\,\eta(t)-s(t)\right)A_0^2\,\frac{a_2}{a_1}\,\cos(b_1+b_2).\label{Jac17}
\end{split}
\end{align}
For the parameter $a_2$, we get
\begin{align}
\begin{split}
\frac{\partial a_2}{\partial t} =&\,\left(q^2\,\eta(t)-s(t)\right)A_0^2a_1\sin(b_1+b_2),\label{Jac18}
\end{split}
\end{align}
and for the parameter $b_2$, the evolution equation is
\begin{align}
\begin{split}
\frac{\partial b_2}{\partial t} = &\,2kq-q^2-\frac{V_0}{2}-s(t)\left(A_0^2+2a_1^2+a_2^2\right)+q^2\,\eta(t)\\
&\,\times\left(A_0^2+4a_1^2\right)+\left(q^2\,\eta(t)-s(t)\right)A_0^2\frac{a_1}{a_2}\cos(b_1+b_2).\label{Jac19}
\end{split}
\end{align}

For simplicity, we may use a variant of the ansatz (\ref{Jac10}) for
which
\begin{align}
\begin{split}
a_1 = a_2 = a, \,\,\,\,\,\,\,\,\,\,\,\,  \mathrm{and}
\,\,\,\,\,\,\,\,\,\,\,\,\,\, b_1+b_2=b.\label{Jac20}
\end{split}
\end{align}
Then the coupled ordinary differential equations for $a(t)$ and
$b(t)$ are
\begin{align}
\begin{split}
\frac{\partial a}{\partial t} = &\,\left(q^2\,\eta(t)-s(t)\right)A_0^2 a\sin(b),\label{Jac21}
\end{split}
\end{align}
and 
\begin{align}
\begin{split}
\frac{\partial b}{\partial t} =&\,-2q^2-V_0+2q^2\,\eta(t)\left(A_0^2+4a^2\right)-2s(t)\left(A_0^2+3a^2\right)\\
&\,+2\left(q^2\,\eta(t)-s(t)\right)A_0^2\cos(b).\label{Jac22}
\end{split}
\end{align}
Using Eq. (\ref{Jac20}), we rewrite the Hamiltonian of the system
given by Eq. (\ref{Jac14}) into a more simple form as
\begin{align}
\begin{split}
H =\pi  \Big\{A_0^2\left(2k^2+V_0+A_0^2 s(t)\right) \\
+A\left[2(k^2+q^2-q^2(A_0^2-A)\eta(t))+V_0\right] \\
+A\left(4A_0^2+\frac{3}{2}A\right)s(t)+2A_0^2A \cos(b)(s(t)
-q^2\,\eta(t)) \Big\}.\label{Jac23}
\end{split}
\end{align}
However, we have found that $A(t)$ and $b(t)$ are canonically
conjugate with respect to an effective Hamiltonian
$H_{\mathrm{eff}}$ (i.e. $\frac{\partial A}{\partial
t}=-\frac{\partial H_{\mathrm{eff}}}{\partial b}$ and
$\frac{\partial b}{\partial t}=\frac{\partial
H_{\mathrm{eff}}}{\partial A}$) which reads
\begin{align}
\begin{split}
H_{\mathrm{eff}}(A) = &\, - \Big[V_0
A+2q^2A+\frac{3}{2}s(t)A^2+2s(t)A_0^2 A
\\
&\, -2 q^2\,\eta(t)(A_0^2+A)A+2(s(t)
\\
&\,  -q^2\,\eta(t))A_0^2 A \cos(b) \Big].\label{Jac24}
\end{split}
\end{align}
In the absence of lattice, this effective Hamiltonian is an exact
integral of motion on the subspace spanned by the ansatz
(\ref{Jac10}). Using $A(t = 0) = 0$ (without loss of generality)
in Eq. (\ref{Jac24}) yields $H_{\mathrm{eff}}^{0} =
H_{\mathrm{eff}}(A(t = 0)) = 0$. Since the Hamiltonian is not
conserved in the presence of OL, the non-conservative part of the
Hamiltonian may transfer an energy, say $E_{\mathrm{com}}$, to the
center of mass of the solitons eventually generated in the case of
MI. Thus we may write $H_{\mathrm{eff}}^{0} - E_{\mathrm{com}}=
H_{\mathrm{eff}}(A)$ and then obtain
\begin{align}
\begin{split}
V_0 A+2q^2A+\frac{3}{2}A^2s(t)+2A_0^2 A s(t)\\
-2A q^2\,\eta(t)(A_0^2+A)+2A_0^2A \cos(b)(s(t) -q^2\,\eta(t))
=E_{\mathrm{com}}. \label{Jac25}
\end{split}
\end{align}
One may easily rewrite Eq. (\ref{Jac21}) in terms of the generalized
momentum $A$. Then eliminating the generalized coordinate $b$
between the resultant equation and Eq. (\ref{Jac25}), we come to the
following energy equation for $A$ :
\begin{align}
\begin{split}
\frac{1}{2}\dot{A}^2+V_{\mathrm{eff}} = 0,&\, \label{Jac26}
\end{split}
\end{align}
where the effective potential reads
\begin{align}
\begin{split}
V_{\mathrm{eff}} = 2A^2q^2[q^2+2s(t)A_0^2-2A_0^2q^2\eta(t)]\\
 +2A^2V_0(q^2+\frac{V_0}{4}+s(t)A_0^2-A_0^2q^2\eta(t))\\
 +A^3[3s(t)(q^2+\frac{V_0}{2}+A_0^2 s(t))]\\
+A^3[q^2\eta(t)\big(4A_0^2q^2\eta(t)-7A_0^2s(t)-4q^2-2V_0\big)]\\
+A^4[
q^2\eta(t)(2q^2\eta(t)-3s(t))+\frac{9}{8}s(t)^2]+V_{\mathrm{eff}}^{0}.
\label{Jac27}
\end{split}
\end{align}
The coefficient $V_{\mathrm{eff}}^{0}$ depends on
$E_{\mathrm{com}}$, $V_s$, $g$ and $\sigma$.

Through the effective potential in Eq. (\ref{Jac27}), we can find
the information about the dynamical properties of the system. For
instance, the evaluation of the curvature of the potential at $A =
0$ may determine whether the dynamics is stable or not. When the
potential is concave, i.e., the second derivative
$\frac{\partial^2 V_{\mathrm{eff}}}{\partial A^2}|_{A=0}$ is
negative, the dynamics is unstable. Otherwise when
$\frac{\partial^2 V_{\mathrm{eff}}}{\partial A^2}|_{A=0}$ is
positive, the potential is convex and the dynamics is stable. We
take advantage of such mathematical evidence to find the MI
criteria of the system. Hence for the dynamics to be unstable, we
should have $\frac{\partial^2 V_{\mathrm{eff}}}{\partial
A^2}|_{A=0}=4\left[q^2(q^2+2s(t)A_0^2+V_0-2A_0^2q^2\eta(t))\right]$
$+4V_0\left[(\frac{V_0}{4}+
s(t)A_0^2-A_0^2q^2\eta(t))\right]+\upsilon_{0}$ negative.  For a
safe domain of parameters, the coefficient
$\upsilon_{0}=\frac{\partial^2 V_{\mathrm{eff}}^{0}}{\partial
A^2}|_{A=0}$ is small and then can be neglected \cite{PLA2013b}.
In such a domain, the condition for exciting the MI reads
\begin{align}
\begin{split}
V_0^2+4(q^2+g\, l(t) A_0^2 - \sigma g\, l(t)^{-1} q^2 A_0^2 )V_0\\
+4q^2(q^2+2g\, l(t) A_0^2-2\sigma g\, l(t)^{-1} q^2 A_0^2 )<0.
\label{Jac28}
\end{split}
\end{align}
To study the instability of the system with respect to the initial
parameters $\alpha$, $V_s$, $g$ and $g_2$, we may rewrite Eq.
(\ref{Jac28}) as
\begin{align}
\begin{split}
V_s^2+4\Big[Q(t)^2 + S(t) A_0^2 - \sigma S(t) Q(t)^2 A_0^2 \Big]V_s\\
+4Q(t)^2\Big[Q(t)^2+2S(t) A_0^2 - 2 \sigma S(t) Q(t)^2 A_0^2 \Big] <
0, \label{Jac29}
\end{split}
\end{align}
where $Q(t)=q(1+4\alpha t^2)^{\frac{1}{2}}$ and
$S(t)=g(1+4\alpha t^2)^{\frac{1}{2}}$, i.e. $Q(t)=q l(t)^{-1}$ and
$S(t)=g l(t)^{-1}$. At the initial time (or in the absence of
harmonic trap), we readily have $Q = q$, $S = g$,
and $V_s = V_0$, and then Eqs. (\ref{Jac28}) and (\ref{Jac29}) are
equivalent. Eq. (\ref{Jac29}) is a time-dependent criterion that
defines the occurrence of MI in BECs with OL and shape-dependent
confinement potential for an atom evolving within the condensates.
In a similar way, Eqs. (\ref{Jac21}) and (\ref{Jac22}) may be
rewritten as follows:
\begin{align}
\begin{split}
\frac{\partial a}{\partial t} = &\,-g\,l^{-1}\left(1-\sigma Q^2 \right)A_0^2 a\sin(b),\\
\frac{\partial b}{\partial t} =&\,-2Q^2-V_s-2g\,l^{-1}\left(1-\sigma
Q^2 \right)A_0^2\\-2g\,l^{-1}&\,\left(3-4\sigma Q^2 \right)a^2
-2g\,l^{-1}\left(1-\sigma Q^2 \right)A_0^2\cos(b).\label{Jac30}
\end{split}
\end{align}
In Eq. (\ref{Jac30}) especially, $Q$ and $l$ are functions of $t$
which is the time in the original dimensionless Eq. (\ref{Jac4}).

Using the conditions in Eqs. (\ref{Jac28}) and  (\ref{Jac29}), we
can find a lot of information on the dynamical instability of the
system with respect to the internal properties of the condensate,
the external perturbation and the external trapping characteristics.
Actually they relate in a very simple way the confinement strength
$\alpha$ of harmonic potential, the strength $V_s$ of optical
potential, the two-body contact interaction term $g$, the two-body
interaction correction term $\sigma$ due to shape-dependent
confinement, and the perturbation wave number $q$. From
Eq.~(\ref{Jac29}), the local growth rate of instability can be
obtained. It is given by the relation
\begin{align}
\begin{split}
\mathrm{Gain}=\Big[-\Big(V_s^2+4(Q^2+S A_0^2- \sigma S Q^2 A_0^2)V_s\\
+4Q^2(Q^2+2S A_0^2-2\sigma S Q^2 A_0^2)\Big)\Big]^{1/2}.
\label{Jac31}
\end{split}
\end{align}

In the particular case where the BEC is not immersed in an OL
potential, that is, $V_s = 0$, we readily obtain from Eq.
(\ref{Jac28}) that $q^2+2g A_0^2(1+4\alpha
t^2)^{-\frac{1}{2}}-2A_0^2q^2g g_2 (1+4\alpha
T^2)^{\frac{1}{2}}<0$, which is exactly the same instability
condition as discussed in Ref. \cite{PRE86}. A similar condition
can also be obtained from Ref. \cite{PLA2013}. Moreover, in the
absence of the HOI and OL potential, i.e. $\sigma=V_s = 0$, we
have $q^2<-2g A_0^2(1+4\alpha t^2)^{-\frac{1}{2}}$, which is the
instability condition obtained in a previous study of MI through
the NLS equation with focusing nonlinearity and quadratic
potential \cite{mitheo2}. From the condition in Eq. (\ref{Jac29}),
one can obtain that the instability diagram in the $q$-$V_s$
plane, for both attractive and repulsive two-body interactions,
has at least three regions bounded by two parabolic lines defined
by $V_s=-2l(t)^{-2}q^2$ and $V_s=-2l(t)^{-2}(1- \sigma
l(t)^{-1}/\sigma_{c})q^2-4g A_0^2 l(t)^{-1}$, with
$\sigma_{c}=1/(2g A_0^2)$.

In what follows, we check in detail the dynamical instability of
the system in both cases of attractive and repulsive two-body
interactions. In calculations, we use $\alpha=0.00633$, $A_0=1.0$,
and $q=0.5$, if not explicitly stated. To obtain the evolution of
the perturbation, we will numerically solve Eq.~(\ref{Jac30}),
with initial conditions $a(0)=0.01$ and $b(0)=0.01$, through a
fourth-order Runge-Kutta scheme. All quantities plotted are
dimensionless.

\begin{figure}[!ht]
\centering
\includegraphics[width=0.50\textwidth]{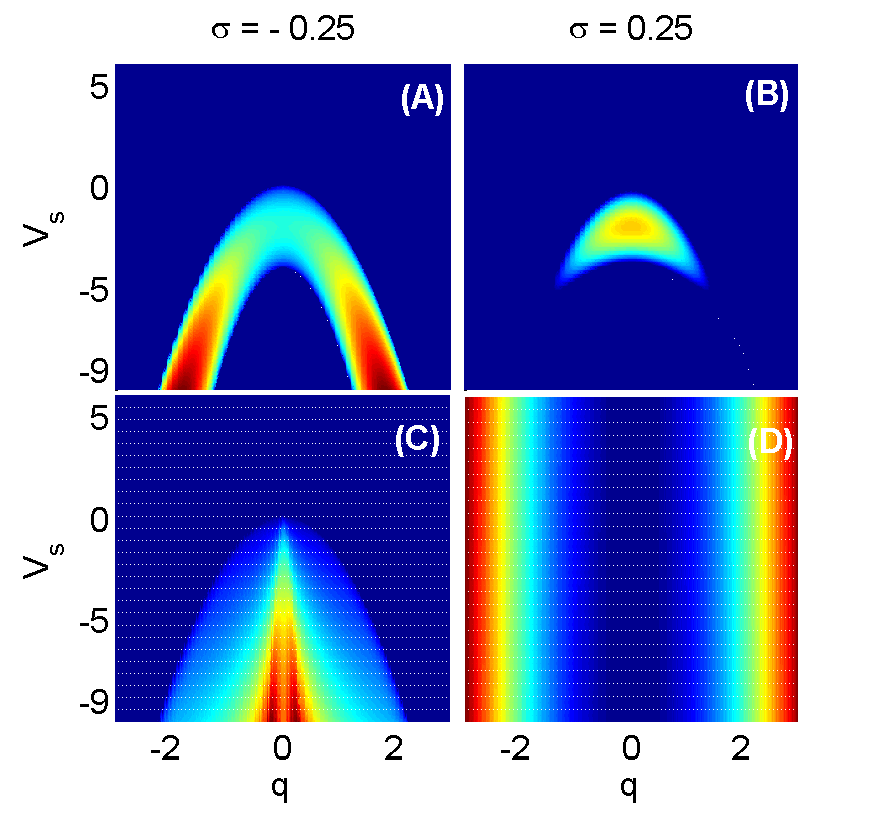}\textbf{}
\caption{Instability diagrams as a function of the OL strength
$V_s$ and wave number $q$ of the excitations, for different values
of the HOI parameter $\sigma$ when \emph{the two-body interaction
is repulsive} ($g=+1$). The dark blue and light colored regions
correspond to the modes for which the system is modulationally
stable and unstable, respectively. The color scale in panels (A)
and (B) indicates the growth rate of the unstable modes at time
$t=0$ for a given $(q, V_s)$. In panels (C) and (D), the color
scale indicates the maximum growth rate of the unstable modes over
time  for a given $(q, V_s)$. The time runs up to $t=9.0$, very
close to the singularity time $\pi/(4\sqrt{\alpha})$ and enough to
allow observing dynamical instability in our system.} \label{fig1}
\end{figure}

\begin{figure}[!ht]
\centering
\includegraphics[width=0.47\textwidth]{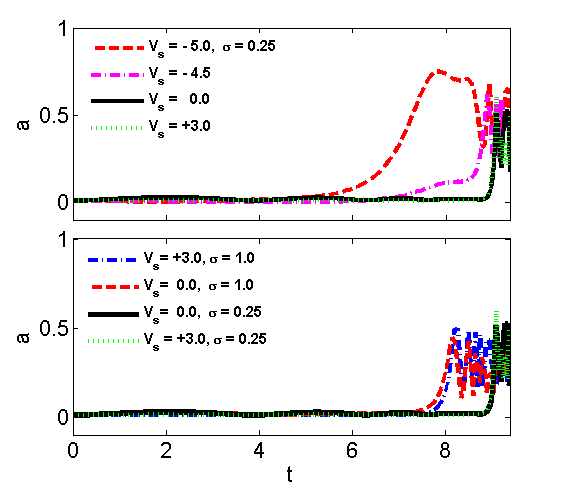}\textbf{}
\caption{The time evolution of the modulational perturbation $a$,
described by Eq. (\ref{Jac30}), for different strengths of the OL
potential, when \emph{the HOI parameter is positive and the
two-body interaction repulsive} ($g=+1$). The initial condition
used in the computation is $a(0)=0.01$. In the top panel,
$\sigma=0.25$. The wave number $q$ is modulationally unstable
(exponential growth) for all OL strengths. In the bottom panel, we
plot the comparative evolution of the perturbation for two values
of the HOI parameter, namely $\sigma=0.25$ and $1.0$, and for
various values of the OL strength. For each value of $\sigma$, the
curves for $V_s=0.0$ and $V_s=3.0$ are almost merged, and then
correspond to the same instability onset time.} \label{fig1a}
\end{figure}

\begin{figure}[!ht]
\centering
\includegraphics[width=0.45\textwidth]{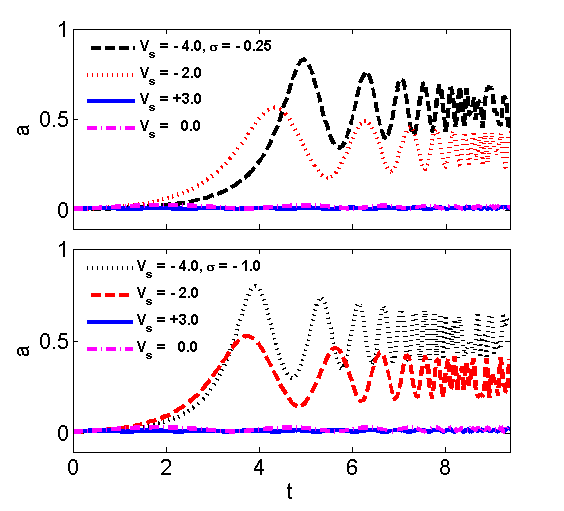}\textbf{}
\caption{The modulational perturbation $a$ as a function of time
for different strengths of the OL potential, when \emph{the HOI
parameter is negative and the two-body interaction repulsive}
($g=+1$). The plots are obtained by solving Eq. (\ref{Jac30}) with
the initial condition $a(0)=0.01$. In the top and bottom panels,
$\sigma=-0.25$, $-1.0$, respectively. Both panels show that the
excitation is modulationally unstable (exponential growth) for OL
strengths $V_s=-4.0$, $-2.0$, and stable for $V_s=0.0$, $+3.0$.
Note that the curves for $V_s=0.0$ and $3.0$ are merged in the
displayed time scale.} \label{fig1b}
\end{figure}

\begin{figure}[!ht]
\centering
\includegraphics[width=0.50\textwidth]{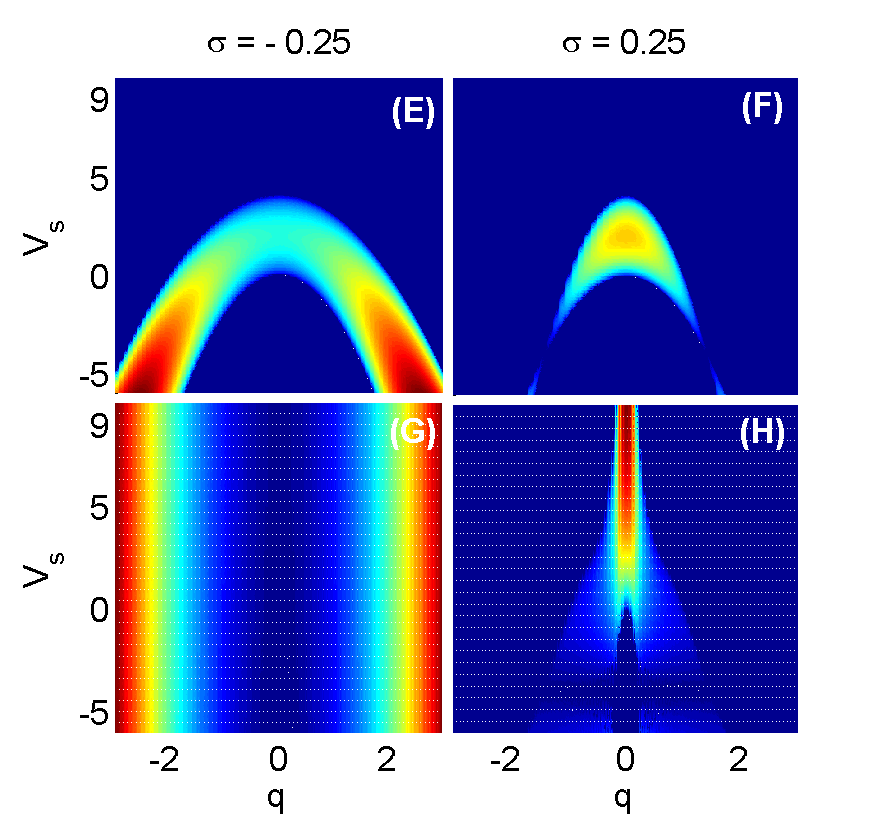}\textbf{}
\caption{Instability diagrams as a function of the OL strength
$V_s$ and wave number $q$ of the excitations, for different values
of the HOI parameter $\sigma$ when \emph{the two-body interaction
is attractive}. The dark blue and light colored regions correspond
to the modes for which the system is modulationally stable and
unstable, respectively. The color scale indicates the growth rate
of the unstable modes for a given $(q, V_s)$. The color scale in
panels (E) and (F) indicates the growth rate of the unstable modes
at time $t=0$ for a given $(q, V_s)$. In panels (G) and (H), the
color scale indicates the maximum growth rate of the unstable
modes over time (for runs up to $t=9.0$) for a given $(q, V_s)$.}
\label{fig2}
\end{figure}

\begin{figure}[!ht]
\centering
\includegraphics[width=0.45\textwidth]{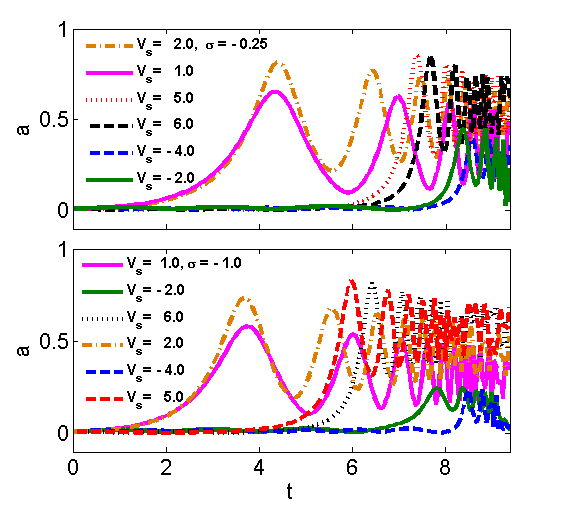}\textbf{}
\caption{The evolution of the modulational perturbation $a$ as a
function of time for different strengths of the OL potential, when
\emph{the HOI parameter is negative and the two-body interaction
attractive} ($g=-1$).  The plots are obtained by solving Eq.
(\ref{Jac30}) with the initial condition $a(0)=0.01$. The wave
number $q$ is modulationally unstable (exponential growth) for all
OL strengths when the HOI parameter is negative.  In the top and
bottom panels, the HOI parameter is $\sigma=-0.25$ and $-1.0$,
respectively.} \label{fig2a}
\end{figure}

\begin{figure}[!ht]
\centering
\includegraphics[width=0.45\textwidth]{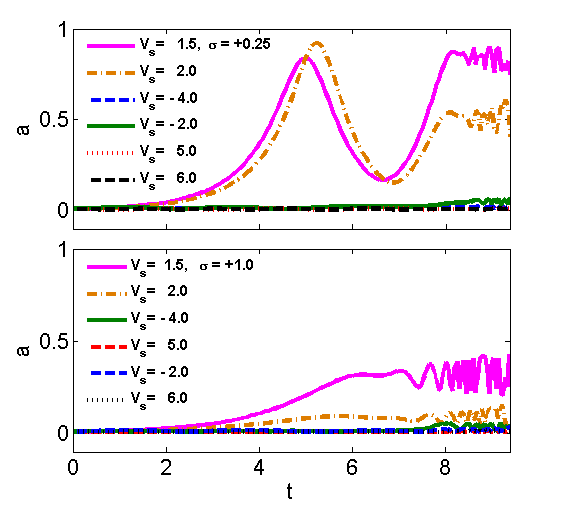}\textbf{}
\caption{The evolution of the modulational perturbation $a$ as a
function of time for different strengths of the OL potential, when
\emph{the HOI parameter is positive and the two-body interaction
attractive} ($g=-1$).  The plots are obtained by solving Eq.
(\ref{Jac30}) with the initial condition $a(0)=0.01$. In the upper
and lower panels, $\sigma=0.25$ and $1.0$, respectively. In the
upper panel, all modes undergo exponential growth from the initial
amplitude. In the lower panel (where the HOI strength is higher),
the plots for the modes with $V_s=-4.0$, $-2.0$, $5.0$ and $6.0$,
almost merged, depict stable oscillations around the initial
amplitude. These modes are all situated in the two initially
stable regions in Fig.~\ref{fig2}(F). However, we realized that
only the modes with $V_s=5.0$ and $6.0$ can keep oscillating at
longer times. The modes with $V_s=-4.0$ and $-2.0$ later develop
exponential growth.} \label{fig2b}
\end{figure}

\subsection{Dynamical instability in the case of two-body repulsion}\label{sec32}

From previous works, we know that dynamical instability does not
arise in a condensate with repulsive two-body interactions ($g>0$)
whatever the excitation wave number is. Here, we predict that the
dynamical instability of a BEC with two-body attraction can
drastically change in the presence of both the OL and the HOI. In
Fig.~\ref{fig1}, we display the instability diagrams of the system
as a function of the OL strength $V_s$ and wave number $q$ of the
excitations for different values of the strength of HOI, $\sigma$,
when the two-body interaction in the condensate is repulsive
($g=+1$). To plot panels (A) and (B), we computed the instability
growth rate of the system at time $t=0$. For panels (C) and (D),
the color scale indicates the maximum over time of the instability
growth rate of the system for runs up to $t=9.0$, which is close
to the analytical limiting time $\pi/(4\sqrt{\alpha})$. Similar
pictures may be obtained by computing the integrated gain instead
of the maximal gain over the same time interval. The dark blue
regions of each panel depict stable modes while the rest of the
domain corresponds to unstable modes.
The shapes of the initial and all-time instability diagrams
displayed in left panels (A) and (C), respectively, are common for
any negative value of the parameter $\sigma$. Besides, while the
all-time instability diagram given in panel (D) is common for all
positive $\sigma$, the shape of the initial instability diagram as
displayed in panel (B) may differ depending whether
$\sigma<\sigma_{c}$, $\sigma=\sigma_{c}$ or $\sigma>\sigma_{c}$,
with $\sigma_{c}=1/(2 g A_0^{2})$ (positive).
The dynamical instability of the BECs with two-body interactions is
strongly determined by the sign of the HOI parameter.

First, we consider the case where $\sigma>0$. From
Figs.~\ref{fig1}(B) and (D), we infer that all modes in the
$q$-$V_s$ plane may eventually lead to dynamical instability.
Moreover at longer times, the instability growth rate in the
system may be almost independent of the OL strength as we have
vertical lines in the color scale in panel (D). Short-wavelength
excitations (with big wave numbers) are completely unstable while
long-wavelength excitations (with small wave numbers) may be
stable for a short time when the perturbation begins. The
occurrence of instability in the system is seriously affected by
the OL strength only for $V_s\leqslant -4 g A_0^{2}$. Meanwhile,
the occurrence of instability is strongly affected by the HOI
parameter. Increasing the parameter $\sigma$ enhances the
instability in the system, and it occurs earlier.
That behavior can be clearly justified through the theoretical
time evolution of the perturbation amplitude in the system. For
this, we numerically solve Eq.~(\ref{Jac30}), with initial
conditions $a(0)=0.01$ and $b(0)=0.01$. The corresponding
evolution of the perturbation amplitude in the system is displayed
in Fig.~\ref{fig1a}. The amplitude of the excitations
exponentially grows for all values of the OL parameter, showing
the instability of the system. In the top panel, for same strength
of HOI ($\sigma=0.25$) the instability onset time ($t \thickapprox
9.1$) is almost identical for $V_s=0.0$ and $3.0$, all greater
than $-4.0\equiv-4 g A_0^{2}$. The instability onset time changes
with the OL parameter for $V_s=-4.5$, $-5.0\leqslant-4.0$; it is,
respectively, $t\thickapprox 8.9$ and $t\thickapprox 7.8$, and so
increases with $|V_s|$. In the bottom panel, the exponential
growth arises around $t=9.1$ for $\sigma=0.25$, and $t=8.1$ for
$\sigma=1.0$ in any of the relevant modes. Then for a bigger
strength of HOI, the instability onset time shortens, showing that
the system becomes more unstable.
Thus the repulsive HOI ($\sigma>0$) is expected to be destabilizing
for a BEC with two-body repulsion. In such a condensate, the OL
strength enhances the dynamical instability for $V_s\leqslant -4 g
A_0^{2}$, and does not affect it for $V_s> -4 g A_0^{2}$.

Secondly, we consider the case where $\sigma \leqslant 0$. From
Figs.~\ref{fig1}(A) and (C), we infer that not all modes may
undergo dynamical instability. The parameter domain in the
$q$-$V_s$ plane has two main regions bounded by the line $V_s=-2
q^2$.
Any excitation with wave number $q$ is expected to be stable for
OL strengths $V_s>-2 q^2$, and unstable for OL strengths $V_s<-2
q^2$. In other words, for $V_s\geqslant 0$ all excitations are
stable. For $V_s<0$ long-wavelength excitations defined by
$q^2<\sqrt{-V_s/2}$ are unstable while short-wavelength
excitations ($q^2>\sqrt{-V_s/2}$) are stable. Within the unstable
domain, the system may become more unstable as the HOI
strengthens. We realized that in such a domain, the instability
gain increases with $|\sigma|$. The unstable bandwidth in the wave
number spectrum enlarges and then more modes become unstable.
We display in Fig.~\ref{fig1b} the theoretical time evolution of
the modulational perturbation to make it clear. In both panels,
the excitation amplitude for the modes with OL strength $V_s=0.0$
and $+3.0$, both greater than $-0.50\equiv-2q^2$, oscillates in
time around its initial value $a(0)=0.01$, which means that the
system is dynamically stable. The excitation amplitude for the
modes with OL strength $V_s=-4.0$ and $-2.0$, all less than
$-2q^2$, exponentially grows in time. Moreover, in the bottom
panel for instance, with the same strength of HOI ($\sigma=-1.0$),
the onset of instability for the unstable OL strengths is
different, notably $t\thickapprox 2.7$ and $3.2$ corresponding to
$V_s=-4.0$ and $-2.0$, respectively.
As we can see from both panels of Fig.~\ref{fig1b}, with the same
strength of OL, the onset of instability for different unstable
HOI strengths is not the same. Notably with $V_s=-4.0$, we get
$t\thickapprox 4.9$ (top panel) and $3.9$ (bottom panel)
corresponding to $\sigma=-0.25$ and $-1.0$, respectively. Thus,
for a bigger strength of HOI (in modulus), the instability onset
time shortens, showing that the system becomes more unstable.
Hence, the occurrence of instability in a condensate with two-body
repulsion depends on the values of the HOI and OL strengths.
Increasing $|\sigma|$ and/or $V_s$ shortens the onset time of the
exponential growth and then enhances the instability in the system
with negative HOI parameter. As shown above, almost similar
results are obtained in the case where the HOI parameter is
positive. With two-body repulsion in the condensate, the
possibility to get modulational stability occurs when $\sigma
\leqslant 0$ and $V_s> -2q^{2}$, with $q$ being the wave number of
the excitation.

\subsection{Dynamical instability in the case of two-body attraction}\label{sec33}

Dynamical instability generally arises in a condensate with
attractive two-body interactions ($g<0$) in a wide range of wave
numbers. In this section, we predict that such a behavior may be
deeply modified in the presence of OL and HOI. To this end, we
portray in Fig.~\ref{fig2} the instability diagrams of the system
as a function of the OL strength and wave number $q$ of the
excitations for different values of the strength of HOI in the
case where the two-body interaction in the condensate is
attractive ($g=-1$). We obtained panels (E) and (F) by computing
the instability growth rate at time $t=0$, and panels (G) and (H)
by computing the maximum instability gain over time for runs up to
$9.0$, close to the limiting time $\pi/(4\sqrt{\alpha})$. The dark
blue regions of the panels may be stable while the rest of the
domain is unstable.
The shapes of the initial and all-time instability diagrams
displayed in right panels (F) and (H), respectively, are common for
any positive value of $\sigma$. Besides, while the all-time
instability diagram given in panel (G) is common for all positive
$\sigma$, the shape of the initial instability diagram as displayed
in panel (E) may differ depending whether  $\sigma<\sigma_{c}$,
$\sigma=\sigma_{c}$ or $\sigma>\sigma_{c}$, with $\sigma_{c}=1/(2 g
A_0^{2})$ (negative).
Similarly to the previous section, we study the dynamical
instability of the BECs with two-body repulsion in two different
regimes of the HOI parameter.

We start with the first regime, where $\sigma<0$. From
Figs.~\ref{fig2}(E) and (G), we infer that all modes may be
dynamically unstable in the relevant time scale as seen in panel
(G). Moreover at longer times, the instability growth rate is
independent of the OL strength as we have vertical lines in the
color scale in that panel. Short-wavelength excitations are
unstable, especially those that lie in the initially unstable
domain while long-wavelength excitations may be stable but for
limited times. In particular, the long-wavelength excitations that
lie in the initially unstable domain, i.e., the light green region
in the middle of the instability diagram in panel (E), are readily
unstable. This suggests that the occurrence of instability is
affected by the OL strength. Meanwhile, the HOI parameter strongly
affects the onset of instability in the system as well. Increasing
$|\sigma|$ enhances the instability in the system.
That feature can be clearly checked through the theoretical time
evolution of the perturbation amplitude in the system. The
corresponding evolution of the perturbation amplitude in the
system is displayed in Fig.~\ref{fig2a}. For calculations, we used
$\alpha=0.00633$ and $q=0.5$. The amplitude of the excitation
exponentially grows for all values of the OL parameter. In the top
panel, for same strength of HOI ($\sigma=-0.25$), the instability
onset time is different for $V_s=-4.0$, $-2.0$, $V_s=1.0$, $2.0$,
and $V_s=5.0$, $6.0$, respectively, picked from the bottom, middle
and top parts of the instability diagram in Fig.~\ref{fig2}(E).
The instability onset times depend on both the OL and HOI
parameters. As we can see in the upper panel of Fig.~\ref{fig2a},
the instability onset time increases in the order $V_s=1.0$,
$2.0$, $5.0$, $6.0$, and $V_s=-2.0$, $-4.0$. Hence, the
instability begins with the modes in the middle region (initially
unstable), then spreads to the modes in the top region (initially
stable), before reaching the modes in the bottom region (initially
stable) depicted in Fig.~\ref{fig2}(E). Within the same region,
the instability onset time weakly increases and the instability
gain slightly decreases when $|V_s|$ increases, except for the
modes in the middle region.
Thus increasing the OL strength softens the instability in the
system for most of the modes. Comparing the upper and lower panels
of Fig.~\ref{fig2a}, it emerges that the instability onset time
shortens when $|\sigma|$ increases. So the attractive HOI
(negative $\sigma$) is destabilizing for a condensate with
two-body attraction, while the OL slightly softens the instability
for initially stable modes and enhances it for initially unstable
modes.

Now let us consider the second regime, where $\sigma \geqslant 0$.
From Figs.~\ref{fig2}(F) and (H), we infer that not all modes may
be dynamically unstable. The parameter domain in the $q$-$V_s$
plane has two main regions as depicted in panel (H). The modes in
the middle region are unstable while the rest of the domain (dark
blue region) may be stable.
Most excitations with wave number $q$ are expected to be stable for
OL strengths $V_s>-4 g A_0^2$. However, some modes satisfying that
condition, and having very small wave number can remain unstable.
For $-4 g A_0^2>V_s>-2/\sigma$, long-wavelength excitations
defined by $q^2<\frac{-V_s-4g A_0^{2}}{2(1-2g A_0^{2})}$ are
unstable while short-wavelength excitations with
$q^2>\frac{-V_s-4g A_0^{2}}{2(1-2g A_0^{2})}$ may be stable. For
$V_s<-2/\sigma$, almost all excitations may be unstable. Within
the unstable domain, the system may become less unstable as the
HOI strengthens.
We display in Fig.~\ref{fig2b} the theoretical time evolution of the
modulational perturbation to clarify that behavior. In both panels,
the excitation amplitude for the modes with OL strengths $V_s=1.5$
and $2.0$ readily undergo exponential growth in time. The growth
onset is delayed in other modes. Hence, considering the instability
diagram in Fig.~\ref{fig2}(F), the instability arises earlier in the
modes from the initially unstable region, later in the lower region
and very late in the upper region. Thus the range where the OL
strength is chosen seriously affects the stability of the
condensate.
Comparing both panels of Fig.~\ref{fig2b}, we obtain that the
repulsive HOI (positive $\sigma$) is stabilizing for a condensate.
In fact, the instability onset is delayed and the instability gain
reduced in all unstable modes when the HOI strength is increased
from $\sigma=0.25$ (upper panel) to $\sigma=1.0$ (lower panel).
For instance, considering the mode with $V_s=1.5$, we obtain the
instability onset times $t\thickapprox 5.0$ (in the upper panel
where $\max(a)\thickapprox 0.9$) and $t\thickapprox 6.0$ (in the
lower panel where $\max(a)\thickapprox 0.4<0.9$).
Moreover, some unstable modes in the upper panel defined by
$V_s>4.0\equiv -4g A_0^{2}$ may even become stable in the lower
panel. However, this stability ends at larger times which increase
with the value of $\sigma$.

Hence, the occurrence of instability in a condensate with two-body
attraction depends on the values of the HOI and OL strengths.
Increasing $\sigma$ enlarges the onset time of the exponential
growth and then softens the instability in the system with
positive HOI parameter. This is in contradiction with the case
above where the HOI parameter is negative. In such case, as we
have previously seen, increasing $|\sigma|$ rather reduces the
onset time of the exponential growth and then enhances the
instability in the system. Meanwhile, increasing $V_s$ delays the
exponential growth and then softens the instability in the system
with positive as well as negative HOI parameter, except for
initially unstable modes where it rather enhances the instability.
With two-body attraction in the condensate, the possibility to get
dynamical stability arises when $\sigma>0$ and $V_s> -4gA_0^{2}$,
with $q$ being the wave number of the excitation. As noted above,
the effective occurrence of this stability requires, in
particular, the systems with sufficiently large values of the HOI
parameter $\sigma$. Moreover, this dynamical stability may vanish
at larger times.

\section{Direct numerical analysis}\label{sec4}
The above analytical results give predictions of the instability
onset in the system. However they are based on the linear stability
analysis of an unperturbed carrier wave. The validity of such an
analysis is limited to amplitudes of perturbation small in
comparison with that of the carrier wave. Moreover, because of the
useful analytical methods, the accessible times are limited to
$\frac{\pi}{4\sqrt{\alpha}} \thickapprox 9.9$ for $\alpha=0.00633$
as used in most calculations. Analytical calculations therefore
cannot tell us, neither the evolution of strongly perturbed wave
nor, the long-time evolution of a modulated extended nonlinear wave.
In order to check the validity of our predictions and go beyond the
limiting time and linearization, we perform direct numerical
integrations of the full GP Eq. (\ref{Jac4}). We used the following
initial configuration \cite{Wamba2008,Wamba2008a},
\begin{align}
\begin{split}
\psi(x,0) =\psi_{\mathrm{TF}} [\phi_0\,\, +\,\varepsilon \, \cos(q
x)],&\, \label{Jac31}
\end{split}
\end{align}
where $\psi_{\mathrm{TF}}=\sqrt{\mathrm{max}[0,\mu-V(x)]}$ is the
background wave function in the Thomas-Fermi approximation, with
$V(x)=\alpha x^2+0.5 \cos(x)^2$. The chemical potential $\mu = 1$
and the strength of magnetic field $\alpha=1/(4\pi)^{2}\approx
0.00633$ are used in all numerical calculations. We have taken
$\varepsilon=0.01$, which is sufficiently small and will not cause
significant variation in the qualitative nature of the results.
The initial amplitude $\phi_0 \equiv A_0=1.0$ is used in all
computations. The boundary condition is periodic, and due to that
we let $q=n\pi \sqrt{\alpha}$. In what follows, we choose $n=2$ to
investigate the effect of both the HOI and the OL in the dynamics
of the corresponding (long-wavelength) excitation. Following the
same spirit as in the analytical part, we consider separately the
cases of two-body repulsion and attraction.

\begin{figure}[!ht]
\centering
\includegraphics[width=0.30\textwidth]{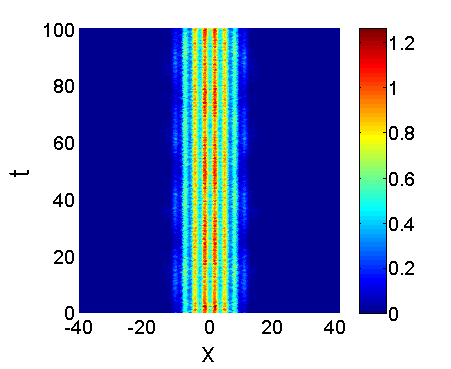}\textbf{(a)}
\includegraphics[width=0.30\textwidth]{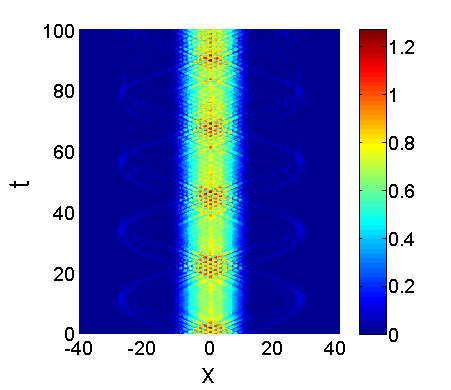}\textbf{(b)}
\includegraphics[width=0.30\textwidth]{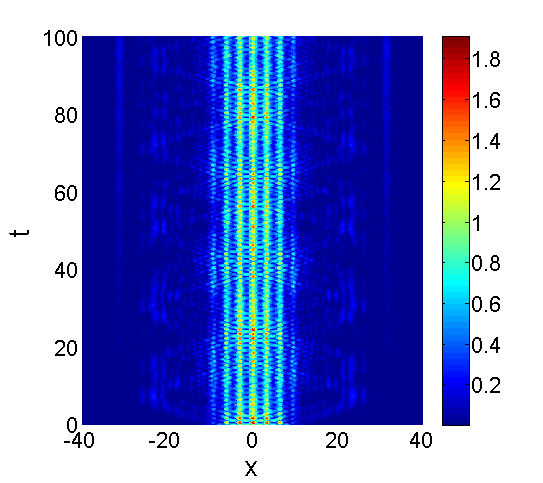}\textbf{(c)}
\caption{Numerical space-time evolution of the wave in the system
with two-body repulsion and attractive HOI for runs up to $t=100$,
and for $g=1.0$, $\sigma=-0.25$, and (a) $V_s=3.0$, (b)
$V_s=-0.25$, (c) $V_s=-4.0$. Mild oscillations in panels (a) and
(b) show that the system is dynamically stable. Panel (c) displays
the oscillatory instability of the system. We note in passing that
we obtained a plot very similar to panel (b) with same parameters
except for $V_s=0.0$ (absence of OL).} \label{num1}
\end{figure}

\begin{figure}[!ht]
\centering
\includegraphics[width=0.35\textwidth]{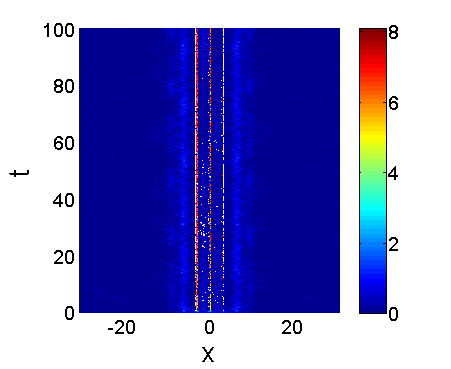}\textbf{(a)}
\includegraphics[width=0.30\textwidth]{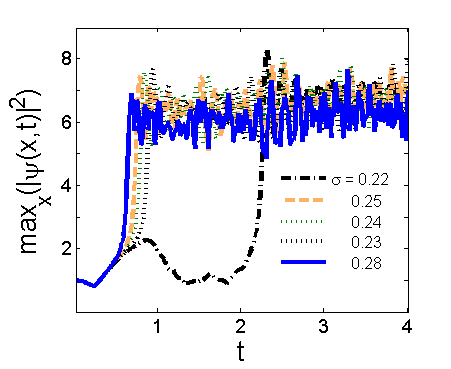}\textbf{(b)}
\caption{Numerical dynamics of the wave in the system with
two-body repulsion and repulsive HOI ($\sigma>0$) for runs up to
$t=100$, and for $g=1.0$ and $V_s>-4.0$. Panel (a) shows the
unstable space-time evolution of the wave for $\sigma=+0.25$ and
$V_s=-2.0$. Panel (b) depicts the comparative evolution of the
wave amplitudes for $V_s=-2.0$ and for $\sigma=+0.22$, $+0.23$,
$+0.25$, $+0.28$. We display the time evolution only up to $t=4.0$
to make visible the separation between the curves, and then the MI
onset times.} \label{num2}
\end{figure}

We begin by launching numerical calculations for the case of BECs,
with \emph{repulsive} two-body interactions ($g=+1$), trapped in
both harmonic and optical fields. As is known from previous works,
such a condensate is dynamically stable for any wave number in the
absence of OL and HOI. Here, we switch on the optical potential,
and check what happens to the stability of the system.

Fig.~\ref{num1} portrays the 3D evolution of the square amplitude
of the wave in the system in time and space for negative HOI
parameter ($\sigma \leqslant 0$). Panels (a) and (b) correspond to
the case where the parameters are picked in the domain predicted
to be stable, i.e., $V_s> -0.5\equiv -2q^{2}$. The difference
between both panels is that in (a) the OL strength is positive and
in (b), it is negative ($0 \geqslant V_s> -0.5$). It should be
noted that the case where the OL is switched off belongs to that
range. Panel (c) corresponds to the case predicted to be unstable.
The plot in Fig.~\ref{num1}(a) shows that the wave amplitude
exhibits mild oscillations both in time and space around its
initial value. Actually the amplitude can go from $1.0$ to roughly
$1.1$ only. The shape as well as density of the condensate are
preserved in time and space. The breathing behavior demonstrates
that the system is completely stable indeed. In Fig.~\ref{num1}(b)
the wave is split into three components. Two very tiny fractions
of the condensate perform regular oscillations in opposite
directions with same frequency and spatial amplitude that goes
beyond the OL. The main fraction exhibits mild oscillations in
time as well and the shape of the condensate is almost preserved
in time and space, demonstrating the stability of the system.
However, in contradiction to the case of panel (a), the
distribution of condensed atoms in the OL wells is not clearly
seen, and a negligible fraction of the condensate can regularly
oscillate in space and time. We suggest that these observations,
which do not alter the stability of the system, are due to the low
height of the interwell barrier of the OL.
The role played by the OL in this stability is not negligible. In
fact, for the same parameters except for $V_s=-4.0$, the system
becomes unstable. We observed the same behavior for $V_s=-3.0<
-0.5$. However, the corresponding instability is oscillatory, as
displayed in Fig.~\ref{num1}(c). The system undergoes irregular
oscillations with higher amplitudes both in time and space. The
maximum square amplitude increases to roughly $1.9$, which is
bigger than $1.3$ obtained in panels (a) and (b) but not big
enough as in usual MI. So we may say that a slight growth rather
than a clear exponential growth is observed in the wave amplitude.
The instability manifests itself not through serious localization,
but in the form of scattering of the wave envelope in space, which
is its salient feature. The many fractions of condensate resulting
from the scattering perform complex oscillations in space and
time. In particular, the condensate fractions loaded in the outer
region of the trap are the most sensitive to oscillatory
instability.
In our context, the sign of the OL strength induces indeed a small
change in the trapping configuration. A positive OL strength
causes the magnetic trap (MT) center to coincide with a well of
the OL, while a negative OL strength causes the MT center to
coincide with an interwell barrier.
Hence trapped BECs with repulsive two-body interactions and
attractive HOI, in the presence of excitations of wave number $q$
and OL of strength $V_s$, exhibit oscillatory instability for
$V_s<-2q^{2}$, i.e., when a high OL interwell barrier exists in
the center of the magnetic trap. This instability, typically
induced by OLs, is characterized by a spatial scattering of the
wave. Such BECs rather exhibit dynamical stability for
$V_s>-2q^{2}$, i.e., when an OL well or a low OL interwell barrier
exists in the MT center. This stability preserves the height and
width of the initial wave.

Let us examine the case of BECs with repulsive two-body
interactions but in the presence of repulsive HOI. When $V_s<
-4.0\equiv -4g A_0^{2}$, the dynamical behavior of the system is
similar to the one plotted in Fig.~\ref{num1}(c). So the BEC
undergoes oscillatory instability. When $V_s> -4.0$, we show in
Fig.~\ref{num2} (a) the 3D space-time evolution of the wave in the
case of positive HOI parameter predicted to be unstable. We
realize that an exponential growth arises in the amplitude of the
wave due to the perturbation. The square amplitude can quickly
increase from $1.0$ to $8.0$, which means that the system is
dynamically unstable. The inner region of the MT is more affected
by dynamical instability than the outer region. However, as we
observed, the instability expands to OL wells in the outer region
of the MT as the OL strength (or the HOI) increases. Thus, in both
cases $V_s< -4.0$ and $V_s> -4.0$, the system is unstable. The
instability is oscillatory for $V_s< -4g A_0^{2}$ and dynamical
for $V_s> -4g A_0^{2}$. In the dynamically unstable parameter
domain, the OL strength causes the instability to expand from the
MT center to the outer region of the MT. Fig.~\ref{num2} (b)
displays the effect of the HOI in the onset of dynamical
instability of the condensate. The OL strength is chosen in the
dynamically unstable domain. The onset time of the exponential
growth shortens when the HOI parameter increases, and the system
becomes more unstable. Thus the positive HOI parameter clearly
appears to be destabilizing for repulsive BECs.

\begin{figure}[!ht]
\centering
\includegraphics[width=0.35\textwidth]{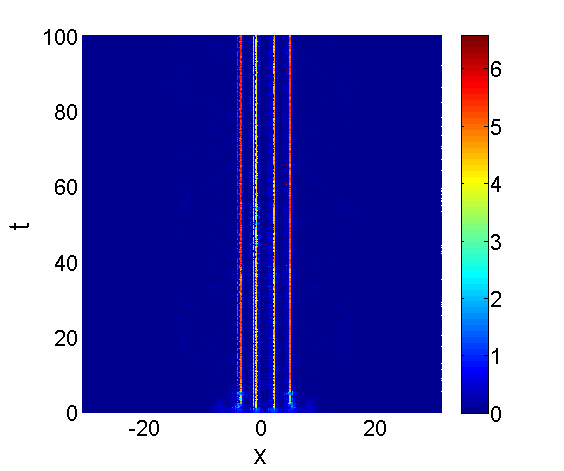}\textbf{}
\caption{Numerical dynamics of the wave in the system with
two-body attraction ($g=-1.0$) and attractive HOI ($\sigma
\leqslant 0$) for runs up to $t=100$. The plot shows the unstable
space-time evolution of the wave for $\sigma=-0.25$ and
$V_s=1.0$.} \label{num3}
\end{figure}

We launch numerical computations for the case of BECs, with
\emph{attractive} two-body interactions ($g=-1$), trapped in both
harmonic and optical fields. As is known from previous works, such
condensate is dynamically unstable for most wave numbers in the
absence of OL and HOI. Here, we show that the stability of the
system dramatically changes when the optical potential is switched
on in the presence of HOI.

Fig.~\ref{num3} portrays the 3D evolution of the wave in time and
space for negative HOI parameter ($\sigma=-0.25$).
We realize that an exponential growth arises in the amplitude of
the wave. As we see the amplitude square can easily go from $1.0$
to about $6.5$, which means that the system is dynamically
unstable. The four middle OL wells which are located in the inner
region of the MT are more affected by MI than the outer wells.
However, as we observed, the instability may expand to OL wells in
the outer region of the MT as the OL strength (or the HOI)
increases.
Thus the attractive two-body and HOI are dynamically destabilizing
for the condensate, and the OL strength cannot help suppressing
such instability.

\begin{figure}[!ht]
\centering
\includegraphics[width=0.35\textwidth]{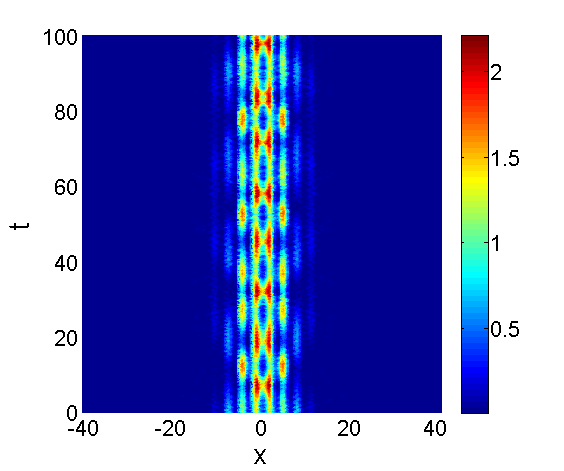}\textbf{(a)}
\includegraphics[width=0.35\textwidth]{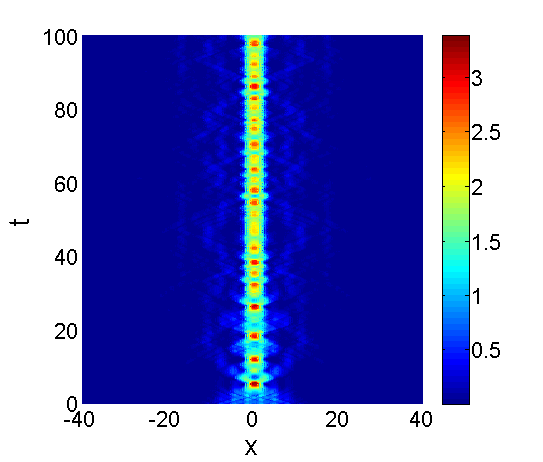}\textbf{(b)}
\caption{Numerical dynamics of the wave in the system with
two-body attraction ($g=-1.0$) and repulsive HOI ($\sigma=+1.0$)
for runs up to $t=100$ and for (a) $V_s>4.0$, and (b) $V_s<4.0$.
The plot in panel (a) depicts the stable space-time evolution of
the wave when $V_s=6.0$. Panel (b) shows the unstable space-time
evolution of the wave when $V_s=1.5$.} \label{num4}
\end{figure}

When the HOI becomes repulsive, the OL strength starts playing an
active part in reducing the instability.
Fig.~\ref{num4} shows the 3D evolution of the wave in time and
space for positive HOI parameter ($\sigma=+1.0$) for two different
values of the OL parameter. Panel (a) shows the case of big OL
strength ($V_s
> 4.0$), while panel (b) corresponds to smaller OL strength ($V_s
< 4.0$).

In the case of deep OL wells as displayed in panel (a), the wave
amplitude slightly increases from $1.0$ to about $1.5$, small
compared to the previous case in Fig.~\ref{num3}. So the
exponential growth does not arise in the amplitude of the wave,
which means that the system is dynamically stable. In all OL
wells, the amplitude exhibits mild oscillations with equal
frequency which do not move the system away from its initial
state. Oscillations in the outer region are out of phase with that
in the inner region. A strong breathing behavior of the wave in
the system is clearly seen.
The OL strength has a perceptible effect on the instability of the
system. As displayed in panel (b), decreasing the depth of OL
wells for instance from $6.0$ to $1.5$ destabilizes the system.
The wave amplitude grows quickly and, simultaneously, a strong
localization of the wave occurs in the center of the OL.
So when the two-body interaction is attractive and the HOI
repulsive, the higher OL strength reduces or prevents the growth
of the wave amplitude in the system, and thus can help suppressing
dynamical instability in the condensate.

Hence, with two-body attraction in the condensate, the system is
dynamically stable when $\sigma>0$ and $V_s> 4$. This stability is
improved for sufficiently large values of the HOI parameter
$\sigma$.

\section{Conclusion}\label{sec5}
In summary, we have analytically and numerically discussed the
dynamical instability of Bose-Einstein condensates with HOI
trapped in parabolic potential embedded in an OL. Through the time
dependent variational approach, we have obtained the set of
ordinary differential equations that govern the evolution of the
perturbation in the system. Moreover, we have established the
criterion that defines the onset of dynamical instability in the
condensate. This work points out the effect of OL depth and HOI
strength in the dynamical stability of the system.

When the two-body interaction is repulsive, the condensates can be
unstable. This occurs whether the HOI is attractive or repulsive.
When the HOI is attractive, the instability is oscillatory,
consisting of the scattering of the wave envelope in space, and
occurs for OL strengths $V_s \leqslant -2 q^2$. When the HOI is
repulsive, the instability is clearly dynamical and manifests
itself through an exponential growth in the wave amplitude. In
this case, increasing the strength of the HOI interaction enhances
the instability in the system.

When the two-body interaction is attractive, the condensates can
be dynamically stable. This occurs when the HOI is repulsive and
the OL wells deep enough. In all OL wells, owing to that
stability, there is neither exponential growth nor scattering of
the wave envelope. The wave amplitude benignly oscillates around
its initial value. The oscillations in the outer region are out of
phase with the oscillations in the inner region of the trap. This
is a clear proof that the system is dynamically stable. Besides,
reducing the depth of OL wells gives rise to the dynamical
instability in the system, clearly characterized by a strong
localization and growth of wave amplitude.

\section*{Acknowledgments}
S.S. thanks the Department of Science and Technology
(DST) for fellowship grant.
KP acknowledges DST, IFCPAR, DST-FCT and CSIR, the Government of
India, for financial support through major projects.
AM thanks the Abdus Salam ICTP for financial support through the
Associateship program.

\section*{References}

\end{document}